\documentclass[12pt]{iopart}
\usepackage[utf8]{inputenc}
\usepackage{graphicx}

\expandafter\let\csname equation*\endcsname\relax
\expandafter\let\csname endequation*\endcsname\relax
\usepackage{amsmath}
\usepackage{amssymb}

\usepackage{xcolor}
\usepackage{url}
\usepackage{booktabs}
\usepackage{multirow}
\usepackage{siunitx}
\usepackage[margin=0.5in]{geometry}

\makeatletter
\def\BState{\State\hskip-\ALG@thistlm}
\let\@fnsymbol\@arabic
\makeatother

\begin{document}

\title{Fast and stable deep-learning predictions of material properties for solid solution alloys}
\author{Massimiliano Lupo Pasini$^1$, Ying Wai Li$^2$, Junqi Yin$^3$, Jiaxin Zhang$^4$, Kipton Barros$^5$ and Markus Eisenbach$^3$}

\address{$^1$ Oak Ridge National Laboratory, Computational Sciences and Engineering Division, Oak Ridge, TN 37831}
\address{$^2$ Los Alamos National Laboratory, Computer, Computational and Statistical Sciences Division, Los Alamos, NM 87545}
\address{$^3$ Oak Ridge National Laboratory, National Center for Computational Sciences, Oak Ridge, TN 37831}
\address{$^4$ Oak Ridge National Laboratory, Computer Science and Mathematics Division, Oak Ridge, TN 37831}
\address{$^5$ Los Alamos National Laboratory, Physics and Chemistry of Materials, Theoretical Division, Los Alamos, NM 87545}
\eads{\mailto{lupopasinim@ornl.gov}, \mailto{eisenbachm@ornl.gov}}

\begin{abstract}
We present a novel deep learning (DL) approach to produce highly accurate predictions of macroscopic physical properties of solid solution binary alloys and magnetic systems. The major idea is to make use of the correlations between different physical properties in alloy systems to improve the prediction accuracy of neural network (NN) models. We use multitasking NN models to simultaneously predict the total energy, charge density and magnetic moment. These physical properties mutually serve as constraints during the training of the multitasking NN, resulting in more reliable DL models because multiple physics properties are correctly learned by a single model. Two binary alloys, copper-gold (CuAu) and iron-platinum (FePt), were studied. Our results show that once the multitasking NN's are trained, they can estimate the material properties for a specific configuration hundreds of times faster than first-principles density functional theory calculations while retaining comparable accuracy. We used a simple measure based on the root-mean-squared errors (RMSE) to quantify the quality of the NN models, and found that the inclusion of charge density and magnetic moment as physical constraints leads to more stable models that exhibit improved accuracy and reduced uncertainty for the energy predictions.

\end{abstract}

{\footnotesize \noindent This manuscript has been authored in part by UT-Battelle, LLC, under contract DE-AC05-00OR22725 with the US Department of Energy (DOE). The US government retains and the publisher, by accepting the article for publication, acknowledges that the US government retains a nonexclusive, paid-up, irrevocable, worldwide license to publish or reproduce the published form of this manuscript, or allow others to do so, for US government purposes. DOE will provide public access to these results of federally sponsored research in accordance with the DOE Public Access Plan (\url{http://energy.gov/downloads/doe-public-access-plan}).}

\normalsize

\section*{Introduction}
Understanding and controlling the properties of materials with different structures is essential for technological progress. However, modeling the behavior of matter at the atomic and molecular level is challenging. Many computational approaches have been developed through the past decades to model and predict the quantum behavior of materials at atomic and molecular scales from first principles, such as density functional theory (DFT) \cite{Hoenberg, Kohn}, quantum Monte Carlo (QMC) \cite{qmc, Hammond} and \textit{ab-initio} molecular dynamics (MD)\cite{car-parrinello, marx}.
To obtain an understanding of the macroscopic physical properties of a material at finite temperature from a thermodynamic perspective, these techniques can be combined with statistical simulations such as classical Monte Carlo methods to reconstruct the probability distribution of the total energy of the system \cite{wang-landau}. This would generally require a large number of samples, hence first principles computations of the total energy and physical observables. However, it quickly becomes computationally formidable as the system size increases, because computing the total energy for each configuration needs numerical integrations in a high-dimensional domain \cite{jolyon}. 

{A specific approach for predicting the total energies of alloys at a lower computational cost is the method of cluster expansion \cite{sanchez, fontaine, levy}. In cluster expansion, the total energy is a summation of the contributions from the interaction energies of different cluster sizes, where the weights or coefficients are determined from a much smaller set of calculations. Practically, the sum is truncated after a certain number of terms, hence the Hamiltonian is a linear combination of a small set of local interactions.}

Another opportunity to reduce the computational time is offered by building inexpensive surrogate NN models to replace expensive first principles calculations \cite{Balabin, Chandrasekaran}. DL models \cite{Chandrasekaran, bhadeshia} are flexible in terms of the range of applications and adjustable computational complexity. {Different from cluster expansion, NN models are able to represent a more general mathematical form and capture non-linear interactions. This ability of NN's makes them a plausible alternative approach to build efficient and robust surrogate models for alloys.} Several studies have already explored how NN models can benefit electronic structure and molecular dynamics calculations \cite{Brockherde, Wang, Sinitskiy, custodio, Ryczko}. {In all cases, the estimation of the total energy needs to be highly accurate to resemble the physics correctly, and stringent requirements on the predictive performance of the NN must be imposed. The use of NN models was restricted to the prediction of one physical quantity using a single NN \cite{LiLi, schleder, meyer}. These NN models can be highly accurate for the quantity it is trained to predict, but they do not make connections to other physical properties of the system to validate the correctness of the predictions. Hence these methods are prone to overfitting for one single physical quantity. Recently, physics-informed machine learning methods have been developed to mitigate the problem. These approaches design the mathematical models or NN architectures that encode and are amenable to known physical properties, geometries or symmetries of a system. Since the physics is incorporated into the machine learning algorithms explicitly, these methods are more transparent and better grounded on physics \cite{purja}.}

This work represents our attempt following the physics-informed design rationale to construct highly accurate, computationally inexpensive, reusable DL models to predict physical properties for solid-state materials, especially binary alloys and magnetic systems. 
Although the total energy is the main macroscopic quantity of interest, other microscopic quantities such as charge distribution and atomic magnetic moment of each atom in the lattice are often objects of interest as well. The joint training method \cite{Caruana} we employ in this work has the advantage that the multitasking NN models can be used to \textit{simultaneously} predict multiple physical quantities.
If these quantities are strongly correlated to each other they can be used as mutual constraints during the joint training. 
Our experiments show that NN models can significantly reduce the computational time to estimate the microscopic physical properties of a material compared to DFT calculations. Moreover, the use of charge density and magnetic moment as physical constraints can improve the accuracy of the predictions of total energy because of the mutual correlations among themselves.

This paper is organized in four sections. Section \ref{physics} describes the physical system, Section \ref{NNmodel} describes the main properties of NN models and training procedure, Section \ref{numerical_section} shows the numerical results for the copper-gold and iron-platinum binary alloys and Section \ref{conclusions} summarizes the results presented and explains possible future directions to continue the research in this field.

\section{Physical system - solid solution binary alloys}
\label{physics}

The physical systems we focus on are solid solution binary alloys, where two constituent atoms are placed on a fixed underlying crystal lattice that does not change with the composition of the alloy. We consider two binary alloy systems. The first case considers the copper-gold (CuAu) alloy with 32 atoms arranged in a $2 \times 2 \times2$ supercell with a face-centered cubic (FCC) structure. The second case considers the iron-platinum (FePt) alloy with 32 atoms arranged in a $2 \times 2 \times4$ supercell with a body-centered cubic (BCC) structure.

For each alloy, 32,000 configurations with different chemical compositions (proportion of atomic species) were chosen to construct the training dataset. More details about dataset construction will follow in Section \ref{dataset_construction}. For each selected configuration, we computed the total energy of the system and the charge density at each atomic position {using the locally self-consistent multiple scattering (LSMS) DFT approach, which exhibits linear sclaing with respect to the number of atoms \cite{eisenbach, lsms, yang}}. We used the LSMS-3 code originated from Oak Ridge National Laboratory \cite{lsms-code}. Since FePt is magnetic, the magnetic moments at all atomic positions were calculated additionally for FePt. In terms of units, the total energy is measured in Rydberg, the atomic charge is measured in electron units and the atomic magnetic moment is measured in Bohr magnetons.

{According to Vegard's law, the chemical composition and the lattice constant of a binary alloy system should be linearly related empirically \cite{vegard, ashcroft}. In this study, however, the DL models were trained for fixed FCC and BCC lattices (and therefore fixed volumes) across different compositions, where the reference lattice constants and volumes were taken as the ones at 50\%/50\% concentration. The reason for doing so is two-fold: to reduce the dimensionality of the input, and to reduce the number of data points (hence DFT calculations) needed for training the models. We stress that although the lattice constant and volume used in our calculations might not be the equilibrium, relaxed ones for a specific chemical composition, the calculations are still correct because the energy is correctly strained to the reference volume. The only limitation is that the DL models will be valid only at this reference lattice constant and volume. }

\section{Deep learning models and training}
\label{NNmodel}

\subsection{Deep feedforward networks}
\label{neural_network}

A deep feedforward network, also called feedforward NN or multilayer perceptron (MLP) \cite{mlp, goodfellow}, is a predictive statistical model to approximate some unknown function $f$ of the form
\[
y=f(x),
\]
where $x\in\mathbb{R}^a$, $y\in \mathbb{R}^b$ and $f:\mathbb{R}^a\rightarrow \mathbb{R}^b$. 
Given a collection of $m$ vectors for $x$ stored in $\mathbf{x}\in \mathbb{R}^{am}$ and the corresponding $m$ vectors of $f(x)$ stored in $\mathbf{y}=f(\mathbf{x})\in \mathbb{R}^{bm}$, feedforward NN's search a mapping that approximates to the best the unknown function $f$. {In this work, $\mathbf{y}$ are the physical quantities of interest, such as the total energy, charge density, or magnetic moment. The input quantities $\mathbf{x}$ are the simplified lattice positions and the atomic species.  Details about the data layout are described in Section \ref{formats}.}

Indeed, a feedforward NN model is composed of many functions stacked together: 
\begin{equation}
\hat{f}(x) = f_\ell(f_{\ell-1}(\ldots f_0(x))),
\label{composition}
\end{equation}
where $\hat{f}:\mathbb{R}^a\rightarrow \mathbb{R}^b$, $f_0:\mathbb{R}^a\rightarrow \mathbb{R}^{k_1}$, $f_\ell:\mathbb{R}^{k_\ell}\rightarrow \mathbb{R}^{b}$ and $f_p:\mathbb{R}^{k_{p}}\rightarrow \mathbb{R}^{k_{p+1}}$ for $p=1,\ldots,\ell-1$.
The proper number $\ell$ will be identified so that the composition in Equation \eqref{composition} resembles the unknown function $f$. The number $\ell$ quantifies the complexity of the composition, and is equal to the number of hidden layers in the NN. Therefore, $f_1$ corresponds to the first hidden layer of the NN, $f_2$ is the second hidden layer and so on. {Each one of the functions $f_i$'s combines the regression coefficients between consecutive hidden layers through nonlinear activation functions that allow deep feedforward networks to create nonlinear relations between input $x$ and output $y$.} The composition in Equation \eqref{composition} is modeled via a directed acyclic graph describing how the functions are composed together.

\subsection{Multitasking learning or joint training}
In this work, we employed a special feedforward NN to predict multiple quantities. This approach is called \textit{multitasking learning} (MTL) or \textit{joint training} \cite{Caruana}. MTL was initially introduced by R.~A.~Caruana to extend the utilization of a NN to simultaneously predict multiple quantities. It provides the capability to exploit the correlations between multiple quantities to improve the accuracy and reduce the uncertainty of the predictions produced by DL models.  The rationale is that predicted quantities that are mutually correlated could act as an inductive bias for each other, leading the learning model to prefer some hypotheses over others. In other words, MTL is expected to retain information that benefit the training of multiple tasks altogether rather than a specific one, because of the domain specific correlation between the multiple quantities to be predicted. 

The strength of an MTL model depends on how strongly the tasks (physical quantities to be predicted) are mutually correlated in a particular application. This type of field specific inductive bias has been defined as \textit{knowledge-based}, for which the training of a quantity can benefit from the information contained in the training signal for other quantities. Indeed, other tasks can provide additional domain knowledge that would otherwise be absent if the quantities were trained independently from each other. Therefore, the mutual correlation between quantities is used to treat every single task as an enrichment of the knowledge available to improve the training of other tasks. This mechanism allows for the explicitly incorporation of physics knowledge to improve model quality.
\begin{figure}[h]
    \centering
    \includegraphics[width=0.15\textwidth]{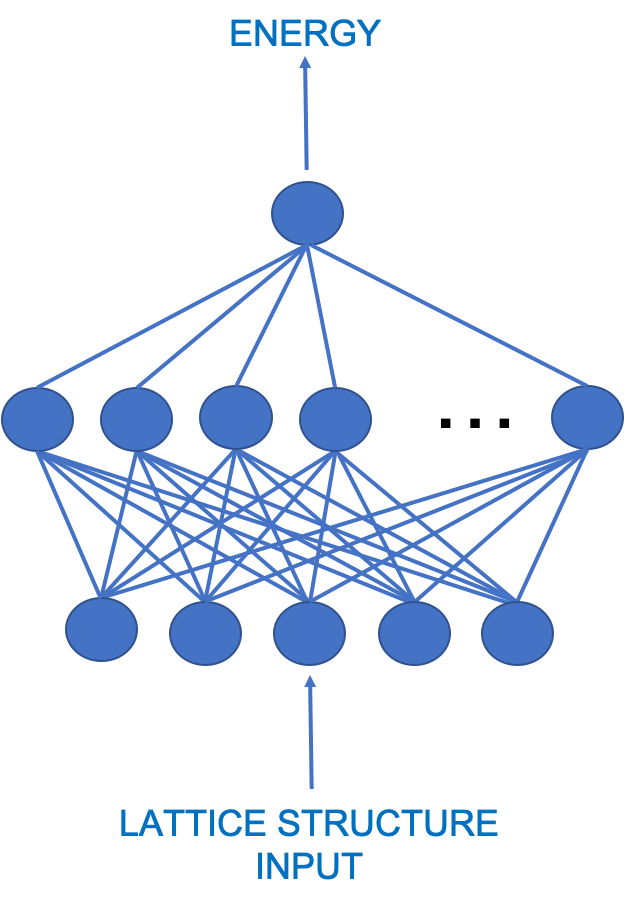}
    \includegraphics[width=0.15\textwidth]{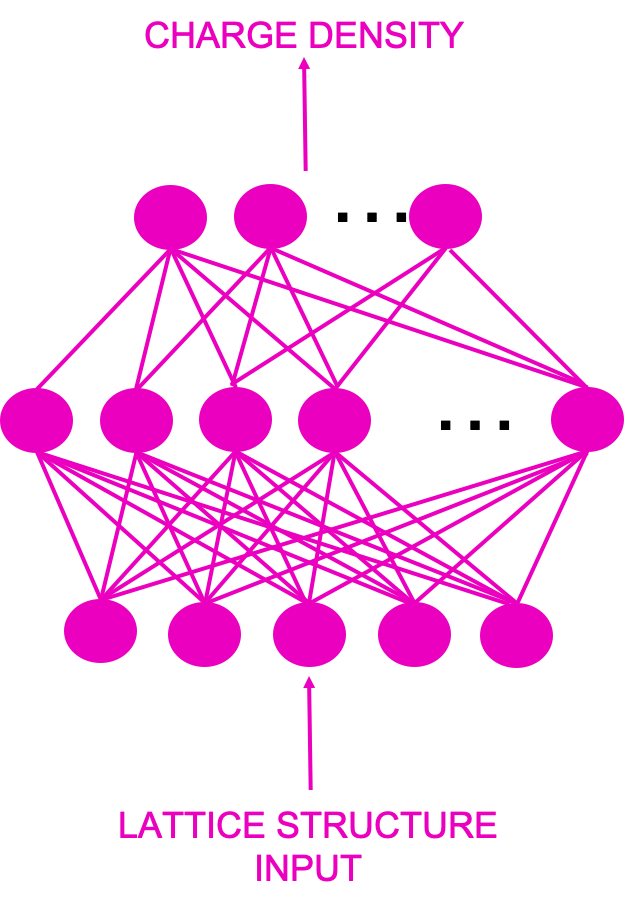}
    \includegraphics[width=0.15\textwidth]{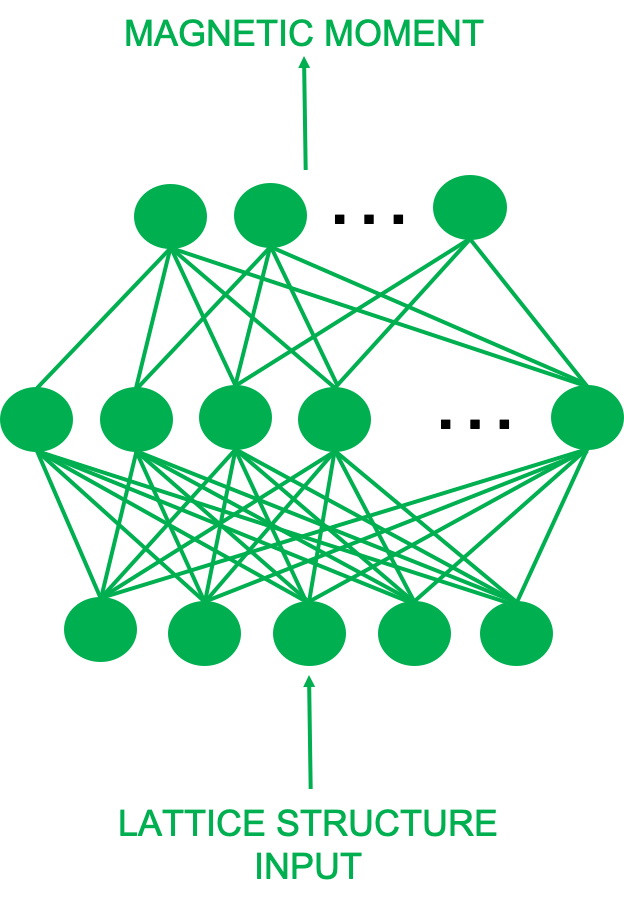}
    \includegraphics[width=0.35\textwidth]{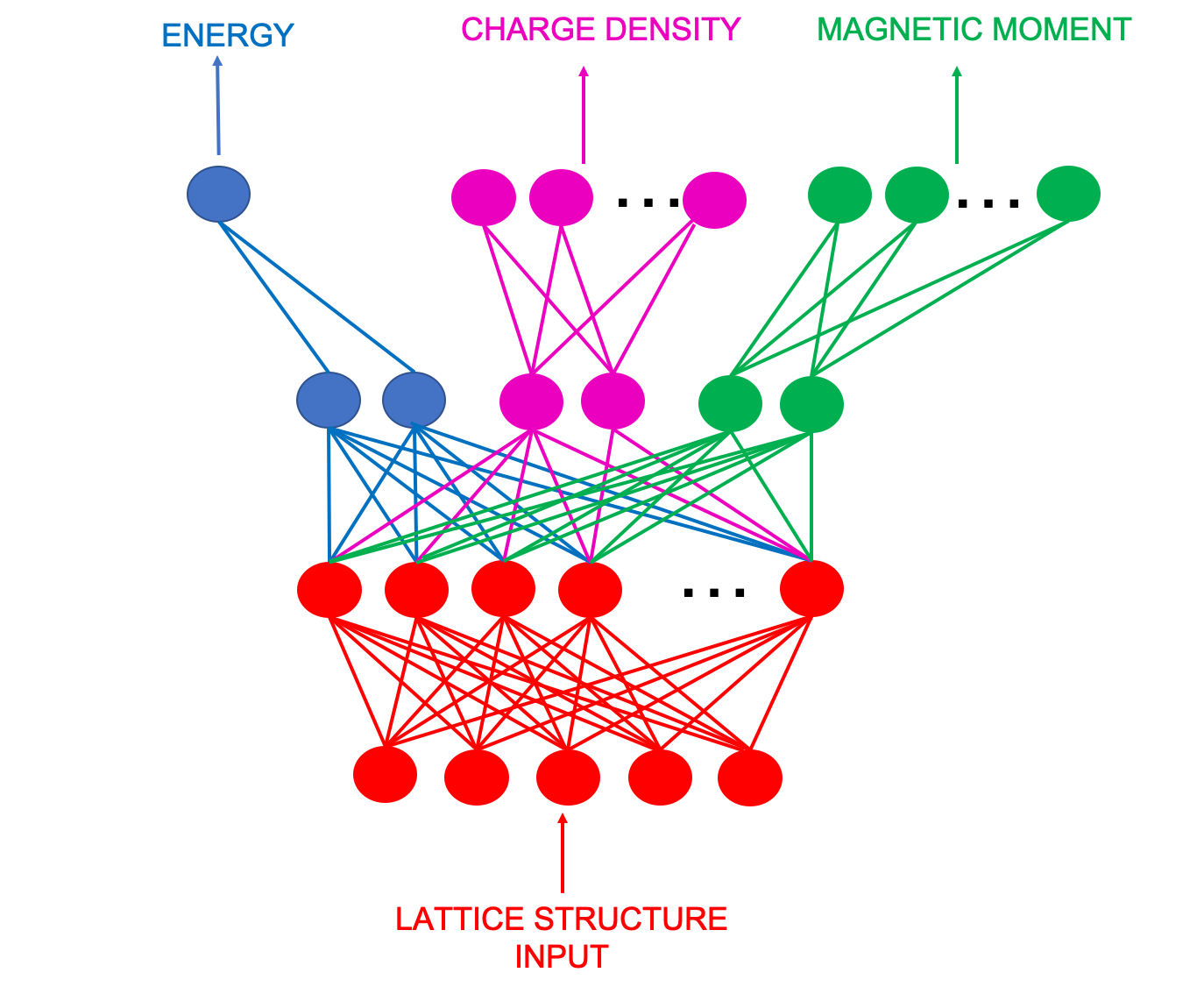}    
    \caption{(Color online.) Architectures of NN models for single-task learning (STL) of total energy, charge density and magnetic moment independently, taking the same inputs (the three NN's on the left), and the architecture of a NN model for multitask learning (MTL) of all three quantities simultaneously taking the same inputs (rightmost).}
    \label{single_and_multiple_tasks}
\end{figure}

Figure \ref{single_and_multiple_tasks} shows the architectures of three NN models to model {three} distinct functions on the same input, and the topology of a NN for multitasking learning to model three related functions defined on the same inputs. The architecture of the NN for MTL is organized so that the hidden layers are shared between all tasks, while keeping several task-specific output layers. This approach is known in literature as \textit{hard parameter sharing}.

\subsection{Regularization}
As any other statistical model used for predictive purposes, feedforward NN models can incur overfitting, which generally refers to numerical artifacts that a predictive model is extremely accurate on the data it was trained to but extremely inaccurate on new samples. Regularization is a technique to alleviate the overfitting problem. This can be expressed as a constrained optimization problem:
\begin{align}
    \begin{cases}
    \underset{\mathbf{w}}{\operatorname{argmin}} \quad \lVert \mathbf{y}_{\textrm{predicted}}(\mathbf{w}) - \mathbf{y} \rVert_2^2\\
    \mathbf{c}(\mathbf{w}) = \mathbf{g} 
    \end{cases},
    \label{strong_constraint}
\end{align}
where $\mathbf{w}\in \mathbb{R}^N$ is the vector of regression coefficients of the model, $\mathbf{c}(\mathbf{w})\in \mathbb{R}^c$ and $\mathbf{g}\in \mathbb{R}^c$ are quantities used to express a constraint. The first equation in Equation (2) represents that the training procedure should find the set of  $\mathbf{w}$ to minimize the difference between the reference values $\mathbf{y}$ and the predicted values $\mathbf{y}_{\textrm{predicted}}$ measured by the $\ell_2$-norm squared. This is also the mean-squared error (MSE) between the predicted and the reference values. The formulation in \eqref{strong_constraint} imposes the constraint $\mathbf{c}(\mathbf{w}) = \mathbf{g}$ in a strong form. The values of $\mathbf{c}(\mathbf{w})$ depend on the model configuration identified by the parameter vector $\mathbf{w}$, whereas the reference values $\mathbf{g}$ are assumed to be given. The constrained optimization problem in \eqref{strong_constraint} can be reformulated so that the constraint is incorporated in the definition of the objective function itself as follows:
%
\begin{equation}
     \underset{\mathbf{w}}{\operatorname{argmin}} \quad \bigg \{  \lVert \mathbf{y}_{\text{predicted}}(\mathbf{w}) - \mathbf{y} \rVert_2^2 + \lambda \lVert \mathbf{c}(\mathbf{w}) - \mathbf{g} \rVert_2^2  \bigg\},
     \label{joint_training}
\end{equation}
which is a global optimization problem. The objective function to be minimized in \eqref{joint_training} interprets the constraint as a penalization term through the \text{penalization multiplier} $\lambda$. This is a weak formulation of the constraint added to the original objective function.

For the multitasking NN in this work, each predicted quantity is associated with a loss function. Therefore, the global objective function to be minimized by the training of the NN is a linear mixing of the individual loss functions. We can view the constraints $\mathbf{c}(\mathbf{w})$ and $\mathbf{g}$ to be some other physical properties to be predicted. In particular, we choose $\mathbf{y}$ to be the total energy of the system, $\mathbf{g}$ as the charge density and/or the magnetic moment, all of them are computed via DFT calculations. Then $\mathbf{y}_{\textrm{predicted}}(\mathbf{w})$ and $\mathbf{c}(\mathbf{w})$ represent their predictions computed via the NN model. 

Formally, let $T$ be the total number of physical quantities, or tasks, we want to predict. A single task identified by index $i$ focuses on reconstructing a function $f_i:\mathbb{R}^{a}\rightarrow \mathbb{R}^{b_i}$ is defined as
\begin{equation}
y_i = f_i(x), \quad i=1,\ldots,T ,
\label{functions}
\end{equation}
where $x \in \mathbb{R}^{a}$, $y_i \in \mathbb{R}^{b_i}$. 
The multitask learning makes use of the correlation between the quantities $y_i$'s, so that the functions $f_i$'s in \eqref{functions} could be replaced by a single function $\displaystyle \hat{f}:\mathbb{R}^a\rightarrow \mathbb{R}^{\sum_i^T b_i}$ that can model all the relations between inputs and outputs as follows: 
\begin{equation}
\begin{bmatrix}
y_1 \\ \vdots \\ y_T \end{bmatrix} = \hat{f}(x).
\label{hat_f}
\end{equation}

The global loss function $\ell_{MTL}:\mathbb{R}^{N_{MTL}}\rightarrow \mathbb{R}^+$ to be minimized in MTL is a linear combination of the loss functions for the single tasks:
\begin{equation}
    \ell_{MTL}(\mathbf{w}_{MTL}) = \sum_{i=1}^T \alpha_i \lVert \mathbf{y}_{\textrm{predict},i}(\mathbf{w}_{MTL}) 
    - \mathbf{y}_i \rVert_2^2,
\label{global_loss}
\end{equation}
where $\mathbf{y}_{\textrm{predict},i}$ is the vector of predictions for the $i^\textrm{th}$ quantity of interest and $\alpha_i$ (for $i=1,\ldots,T$) are the mixing weights for the loss functions associated with each single quantity. 
{ The value of the $\alpha_i$'s in Equation \ref{global_loss} are hyperparameters of the surrogate model and thus can be tuned. In our examples, we assigned an equal weight to each property being predicted, because the data on which the model was trained was properly standardized. However, the joint training and the definition of the loss function enable one to accordingly account for this by adequately modifying the value of the $\alpha_i$'s if it is known that one property dominates over the others. }

As we mentioned earlier, the multiple quantities in MTL can be interpreted as a mutual inductive bias because the MSE of a single quantity acts as a regularizer with respect to the loss functions of other quantities. For a fair comparison, we do not use regularizers for the individual single-task training in this work, so as to examine the benefit of using other tasks as a mutual regularizer.

\subsection{Neural network architectures and hyperparameters for this study}

In our CuAu test case, the single-tasking NN models identified by the hyperparameter search has two hidden layers and each hidden layer contains 200 nodes. The multitasking NN model identified by the hyperparameter search is made of three hidden layers instead; the first two hidden layers have 200 nodes that are shared across all the tasks, whereas the last hidden layer is task-specific (see Figure \ref{single_and_multiple_tasks}). The total number of nodes in the task-specific hidden layer is still 200, but they are equally split across the two tasks to predict the total energy and charge density. The rectified linear unit `\texttt{relu}' is used at each hidden layer as the activation function, whereas the output layer has no activation function as recommended by standard practices in DL when NN models are used for regression problems. 

The loss functions associated with each predicted quantity are linearly mixed to create a global objective function that is minimized through the joint training. The mixing weights $\alpha_i$'s to define the global loss function in Equation \ref{global_loss} are set to 1.0 for each quantity predicted because the input and output data underwent a standardization pre-/post-processing. 

For the FePt test case, the single-tasking NN selected by the hyperparameter search comprises of two hidden layers and each hidden layer contains 300 nodes. The NN for multitasking identified by the hyperparameter search is made of three hidden layers. The first two hidden layers have 300 nodes each and they are shared across all the tasks, whereas the last hidden layer is task-specific. Similar to the CuAu case, the 300 nodes in the task-specific hidden layer are equally split among the three tasks that respectively predict the total energy, charge density and magnetic moment. The presence of the magnetic moment as an additional quantity increases the number of ways that the physical quantities can be combined for the joint training. Also in this case, the mixing coefficients $\alpha_i$'s for the loss functions are set to 1.0 for each quantity predicted. 

The NN architectures described above were identified by using a hyperparameter optimization procedure we devised \cite{lupopasini}. This procedure was designed to identify a NN that attains a desired performance with the minimal structural complexity, specifically, by minimizing the number of hidden layers of the NN. 
The algorithm starts from one hidden layer, and iteratively enriches the architecture of the NN by expanding the number of hidden layers in an adaptive fashion. At each iteration, a random search is performed to identify the number of nodes per layer and the activation function. In order to reduce the granularity of the hyperparameter search in each hidden layer, the possible number of nodes per layer is restricted to multiples of fifty. The set of nonlinear activation functions explored are the hyperbolic tangent, the sigmoid and \texttt{`relu'}.

The selected values of hyperparameters in each layer are transferred to the next iteration, so that the new neural networks are built by reusing the hyperparameter selection already performed for previous hidden layers during earlier iterations.
Different from grid search and random search commonly used in hyperparameter optimization, our algorithm performs random search restricted to one hidden layer at each iteration. This reduces the dimensionality of the hyperparameter space to explore.

This iterative procedure stops when the validation loss of a neural network is smaller than the preset threshold of $10^{-3}$ for the stopping criterion, which is defined by the global loss function in Equation \ref{global_loss} multiplied by a factor that is inversely proportional to the number of nodes in the layer that is being optimized. By doing so, wider neural networks (with more nodes per layer) are penalized over thinner neural networks (with fewer nodes per layer), which favors NN's with minimal complexity. This penalization over wider NN's is also justified by the fact that complex relations between inputs and outputs are better described by expanding the NN's depth rather than the width.

\subsection{Data formats}
\label{formats}

The CuAu binary alloy is a non-magnetizable alloy and each data point in the dataset provides the information about the atomic positions on a lattice with the total energy and the charge density. The FePt alloy has magnetic properties; each data point in the dataset contains information about the atomic positions on a lattice, the charge density and the magnetic moment as well.

\begin{figure}[h]
    \centering
    \includegraphics[width=0.45\textwidth]{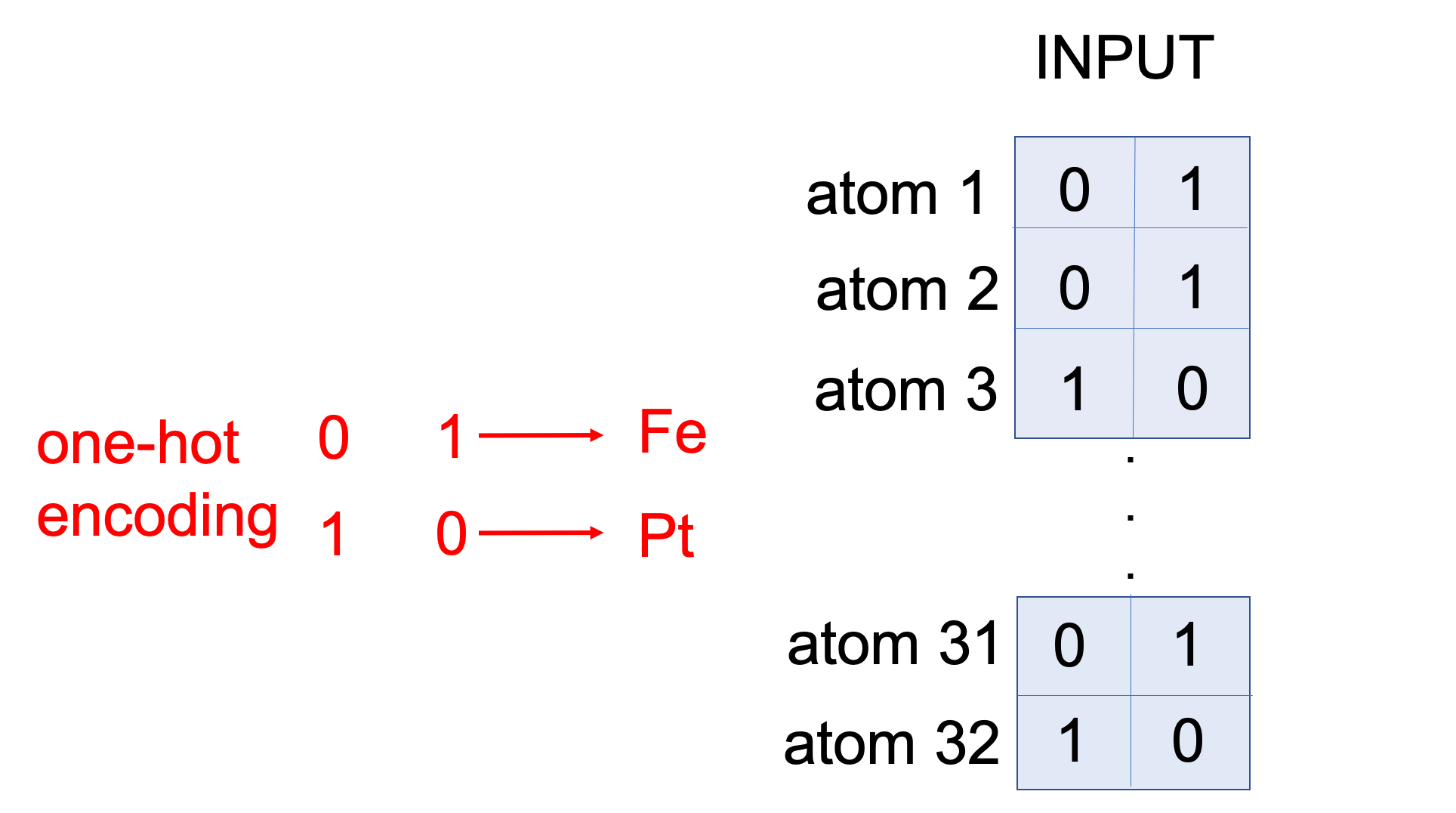}
    \includegraphics[width=0.47\textwidth]{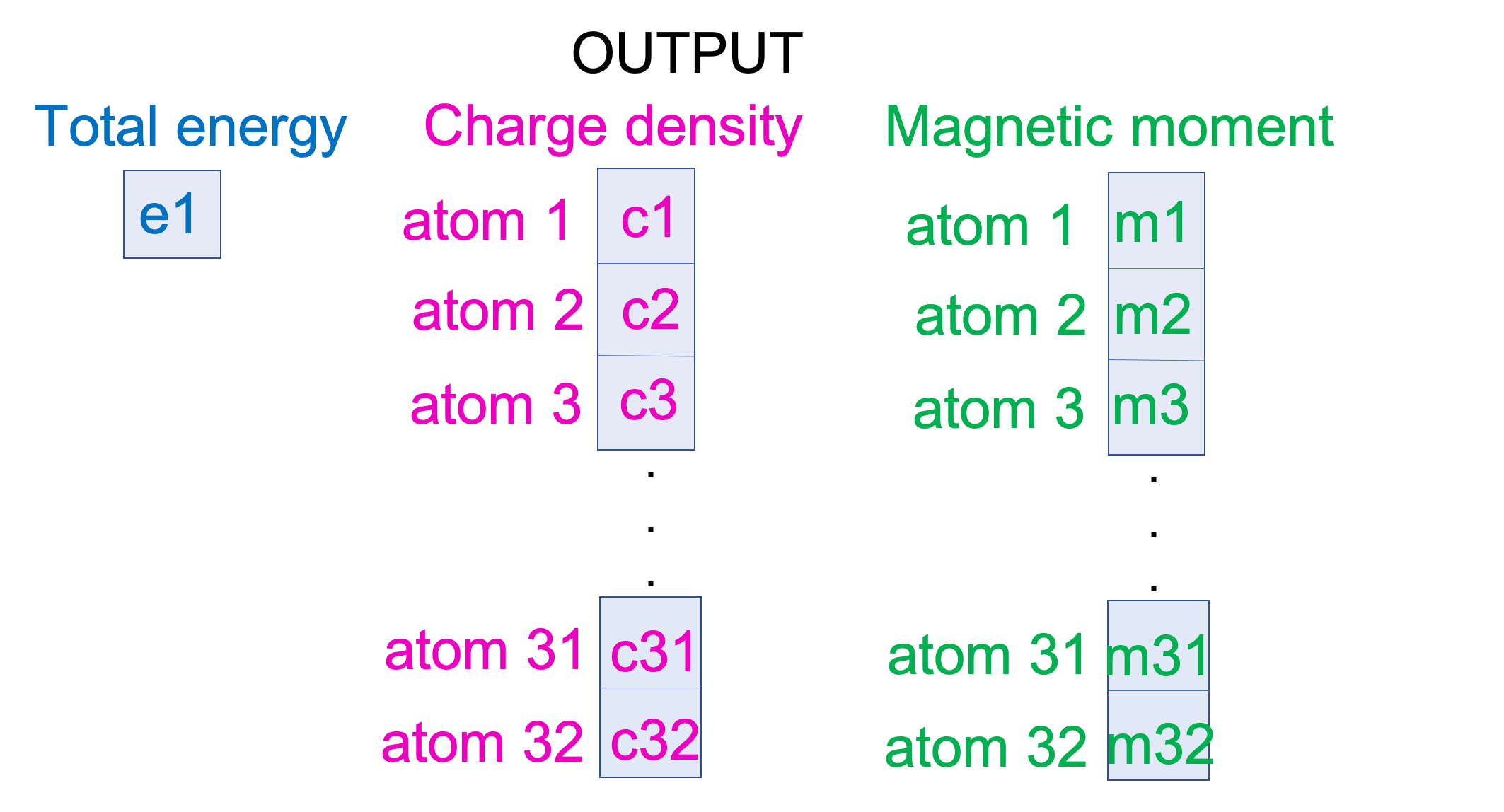}
    \caption{Input data (left) and output data (right) formats for each FePt configuration.}
    \label{input_output_illustration}
\end{figure}

{Figure \ref{input_output_illustration} illustrates the input and output formats employed in the training process. Each input data point is represented as a $32\times 2$ matrix, where each row provides a one-hot-encoding representation of the atomic species located at a specific lattice point, the total energy is represented as a scalar for every configuration, whereas charge density and magnetic moment are represented as one-dimensional vectors with 32 entries for each configuration.}

\begin{figure}[h]
    \centering
    \includegraphics[width=0.2\textwidth]{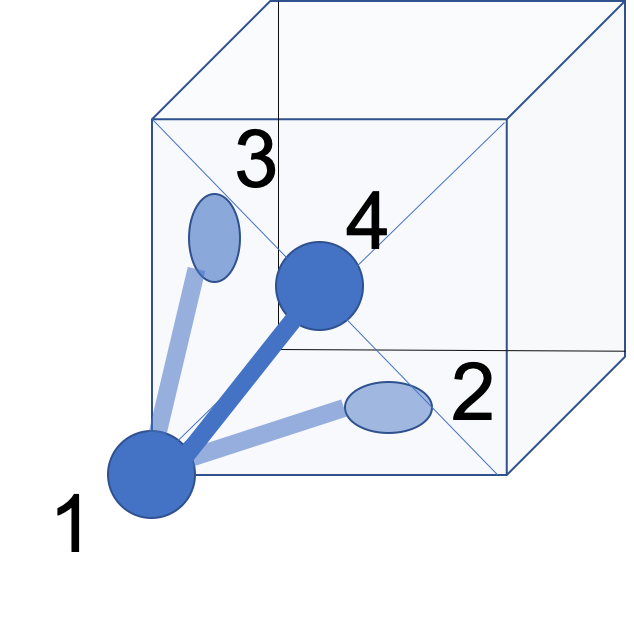}\qquad 
    \includegraphics[width=0.185\textwidth]{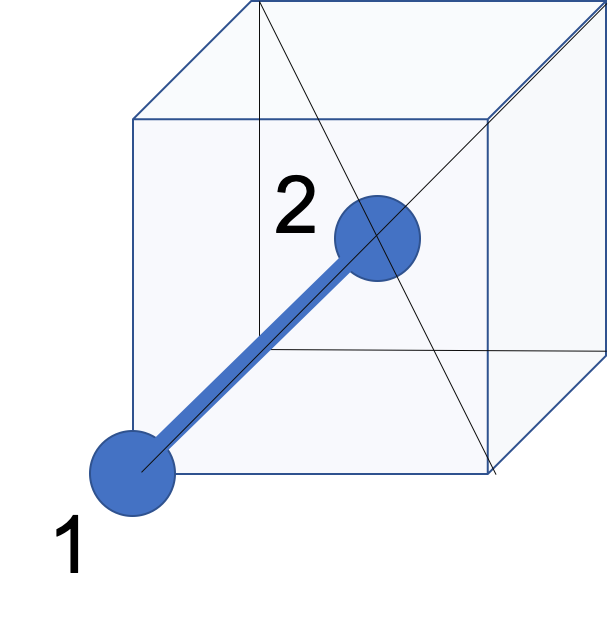}
    \caption{A face-centered cubic (FCC) unit cell for CuAu (left) and a body-centered cubic (BCC) unit cell for FePt (right). The local numbering of atoms in the unit cell for each structure is shown. }
    \label{lattice_structure}
\end{figure}

To reduce the input dimensionality, the two-dimensional, one-hot-encoding input data indicating  the atomic species are arranged in order from atom 1 to atom 32. Since the lattice points of a crystal structure are fixed, there is no need for the three-dimensional coordinates of the atomic positions.  Figure \ref{lattice_structure} shows the local numbering of atoms inside a FCC or a BCC unit cell. For the 32 atoms in CuAu arranged in $2 \times 2 \times2$ FCC unit cells, the atoms are numbered starting from the ($x = 0, y = 0, z = 0$) unit cell. The counting traverses through the unit cells in the $x$-direction first, then the $y$-direction, and the $z$-direction last. For example, atoms 1 to 4 are from unit cell (0, 0, 0), atoms 5 to 8 are from unit cell (1, 0, 0), atoms 9 to 12 are from unit cell (0, 1, 0), and so on. Within a unit cell, the order of counting follows the local numbering of atoms. The numbering of atoms for FePt follows the same manner, with a difference that the 32 atoms are arranged in a supercell having $2 \times 2 \times4$ BCC unit cells, each unit cell only has two atoms.

{
\subsection{Data pre- and post-processing}
The inputs and outputs are respectively standardized (normalized) across all data points for each quantity, such that each quantity has a zero mean and unit standard deviation. Because different physical quantities have different units and different orders of magnitude, the reason for standardization is to avoid artifacts in the {regression coefficients}. After the model is trained, the predicted quantities are rescaled back to restore their actual values and units.
}

\subsection{Dataset construction}
\label{dataset_construction}

\begin{figure}[h]
    \centering
    \includegraphics[width=0.5\textwidth]{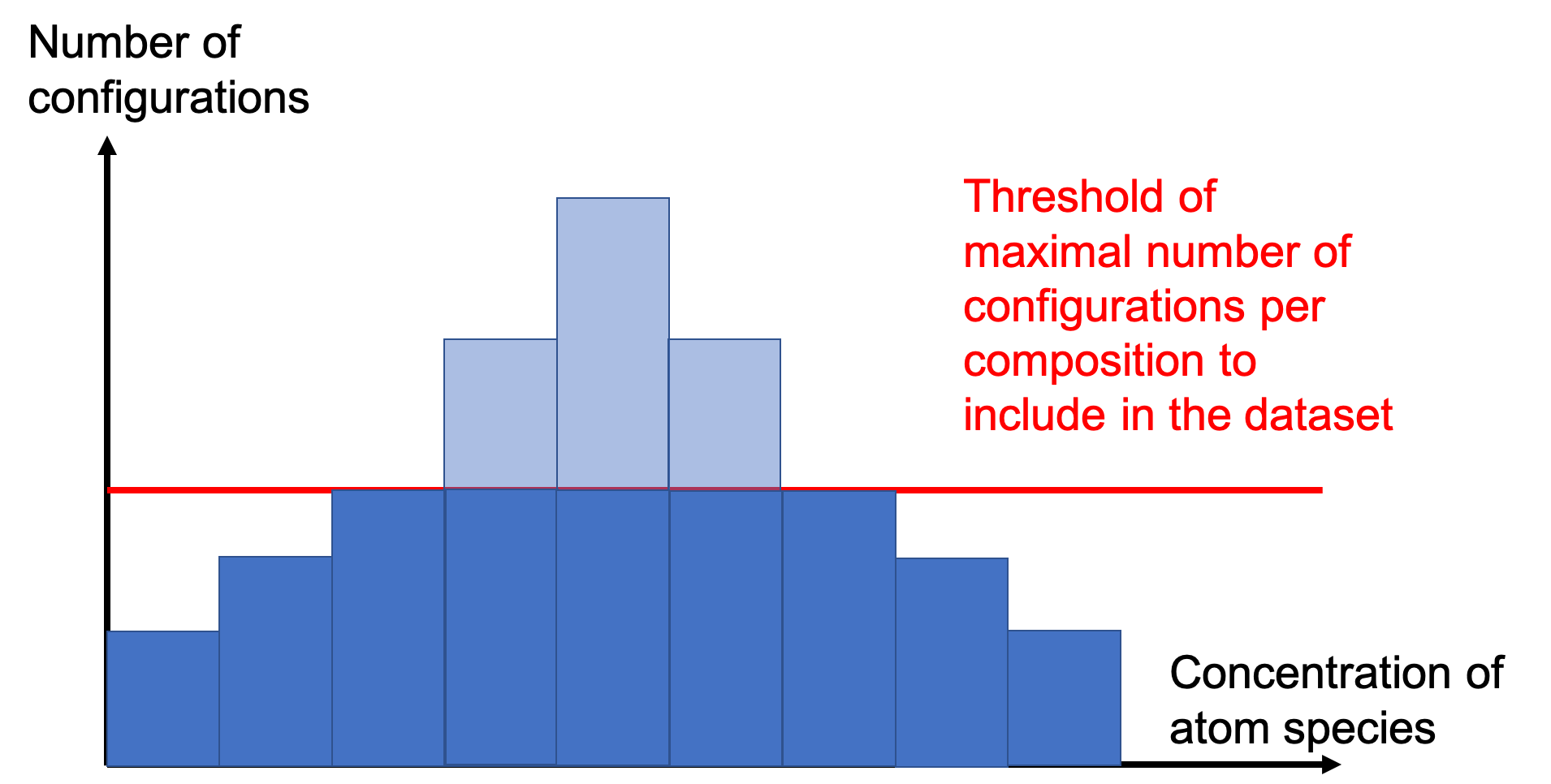}
    \caption{Illustration of the threshold cut to decide the number of configurations for each composition to include in the dataset. If a specific composition has a total number of configurations that is lower than the threshold, then all the configurations for that composition are included in the dataset. If the total number of configurations for a specific composition is higher than the threshold number, a subset randomly selected from those configurations is included in the dataset.}
    \label{configuration_selection}
\end{figure}

The dataset for each alloy comprises a number of lattice configurations for each chemical composition. 
{For both test cases, the dataset is created by extracting 32,000 configurations out of the $2^{32}$ available. The selection of the configurations is performed to ensure that every composition is adequately represented in the dataset. In particular, we select 1,000 as the maximal number of configurations for a fixed composition. If the number of configurations for a specific composition is less than 1,000, then all those configurations are included in the dataset. Otherwise if the number of configurations for a specific composition is more than 1,000, the 1,000 configurations to include in the dataset are randomly selected among all the possible configurations. Moreover, the splitting between training, validation and and testing set is performed at the level of each composition, to ensure that all the compositions are adequately represented in both training and testing portions of the dataset.} {This selection process is illustrated in Figure \ref{configuration_selection}.}

\subsection{Training details}

The {regression coefficients} of the NN are optimized with the Adam method \cite{adam} with an initial learning rate equal to 0.0001. A total number of 30,000 epochs was performed. The batch size for each step of an epoch is 10\% of the training set. {Although early stopping is a common practice in DL to interrupt the training when further epochs would be unlikely to further reduce the loss function, we do not adopt this procedure. It is because we intended to set up a fair experiment to show that a better uncertainty reduction can be achieved by using a multitasking NN instead of single tasking NN for the same fixed computation workload. An early stopping criterion may cause some models to stop the training earlier than others, which would make it difficult to substantiate our claim.} 

{A k-fold cross-validation with $k = 60$ has been adopted to generate 60 different testing and training sets from the 32,000 available data points. To guarantee that all these $k = 60$ training-test splittings have an equal representation of the compositions in both the training and testing portion of the dataset, a k-fold cross validation is first performed within \textit{each} composition, then the portions are combined to generate the global training-validation-test splitting that spans all the compositions.
We then trained a NN for each of the 60 different training sets, hence a total of 60 NN models were trained for each training method. The training set for each of the NN represents 81\% of the total dataset, the validation set represents 9\% of the total dataset and the remaining 10\% is used to test the predictive performance of each DL model.}

\section{Numerical results and discussions}
\label{numerical_section}

We now present the numerical results of training surrogate DL models to estimate physical properties (e.g. total energy, charge density and magnetic moment) for binary alloys CuAu and FePt. We first analyze the training and validation loss functions to verify that NNs were trained correctly. We then assess the computational time reduction obtained by using the DL surrogate models, and the advantage of using the joint training method over single-task training in terms of accuracy and uncertainty reduction.

\subsection{Training and validation losses}
\begin{figure}[h]
    \centering
    \includegraphics[width=0.45\textwidth]{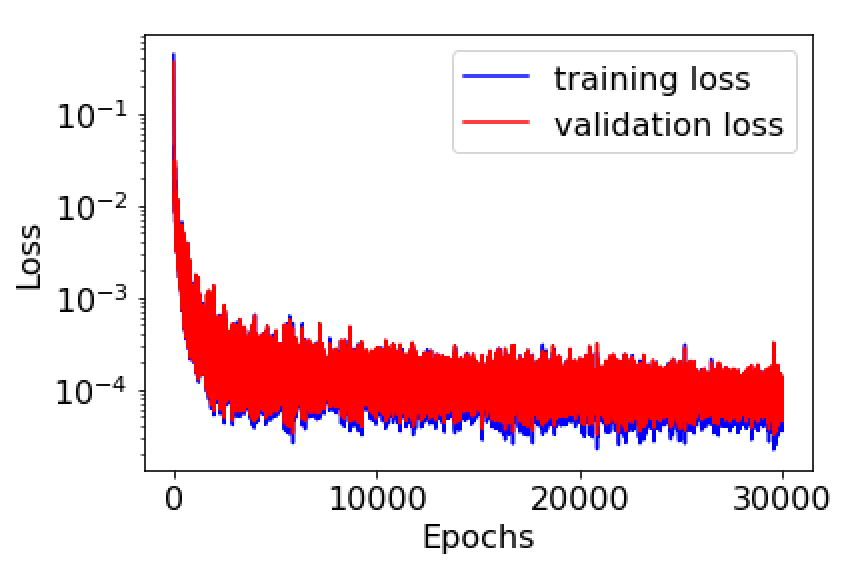}
    \includegraphics[width=0.45\textwidth]{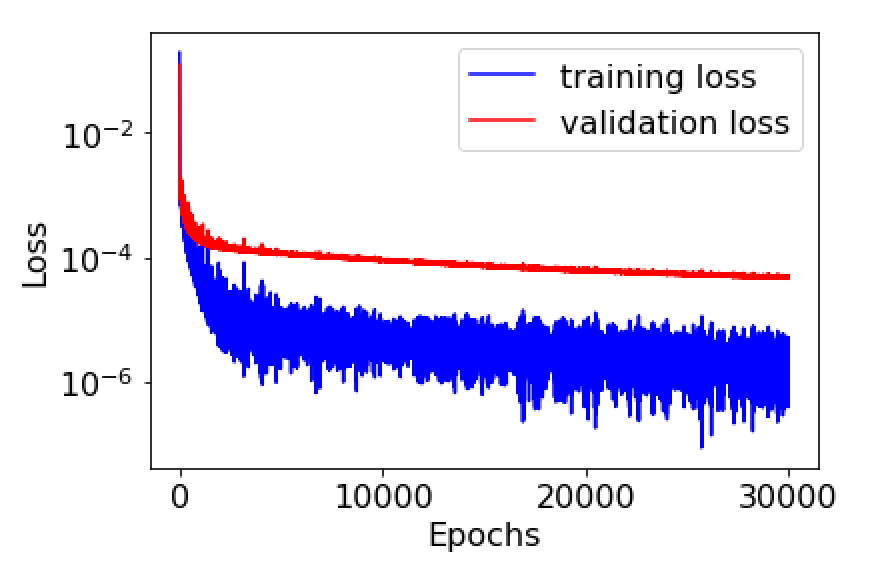}
    \caption{CuAu test case - training loss (blue line) and validation loss (red line) as a function of the epochs during the training of a multitasking NN that predicts both the total energy and charge density (left), and same quantities for the single-tasking NN that predicts the total energy only (right).}
    \label{history_loss_test1}
\end{figure}

{Figure \ref{history_loss_test1} shows the training and validation loss functions for the multitasking NN that simultaneously predict total energy and charge density for CuAu, in comparison with the same quantities for the single-tasking neural networks that predicts the total energy alone. The history of the training and validation loss function shows that overfitting is not happening.} 

 \begin{figure}[h]
    \centering
    \includegraphics[width=0.45\textwidth]{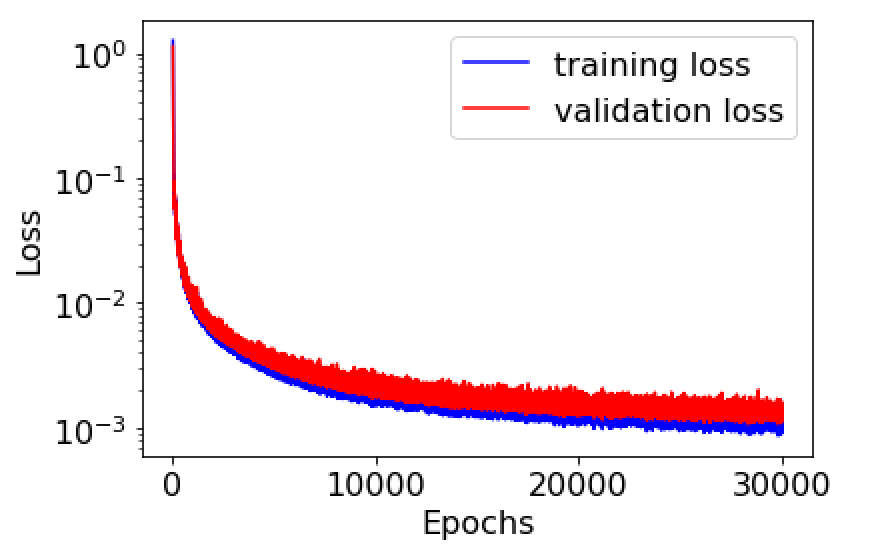}
    \includegraphics[width=0.45\textwidth]{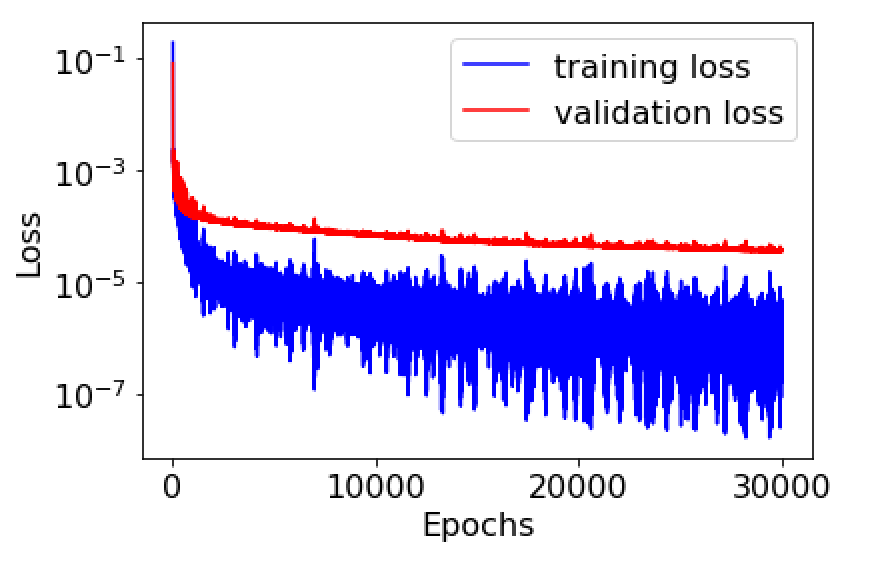}
    \caption{FePt test case - training loss (blue line) and validation loss (red line) as a function of the epochs during the training of a multitasking NN that predicts total energy, charge density and magnetic moment (left), and same quantities for the single-tasking NN that predicts the total energy only (right).}
    \label{history_loss_test2}
\end{figure}

{For FePt, the training and validation loss functions, as well as the accuracy of the predictions also behave similarly as the CuAu case. Figure \ref{history_loss_test2} shows the training and validation loss function for the multitasking NN that simultaneously predict total energy, charge density and magnetic moment, in comparison with the training and validation loss function for the single-tasking neural networks that predicts the total energy alone. Again, the training and validation loss functions show that overfitting is not happening.}

\subsection{Comparison between computational times for first principles calculations and DL models}

In this section we compare the time needed to estimate the total energy for a lattice configuration using first-principles calculations, single-tasking NN models and multitasking NN models for CuAu and FePt.  The output of DFT calculations is considered as the exact reference that the DL model has to reconstruct. Therefore, the predictive performance of a DL model is tested by measuring the departure of quantities predicted by the DL models from the results produced by DFT calculations.

The DFT calculations to generate the dataset to train the NN models were performed with the LSMS-3 code on the Titan supercomputer at Oak Ridge National Laboratory (ORNL), each calculation used 8 Titan nodes (each Titan node had 1 NVIDIA K20X GPU) for a hybrid MPI-CUDA parallelization. For more details about the hardware specifics of Titan we refer to \cite{titan}. Because ORNL's Titan supercomputer has been decommissioned as of this writing, we present here the computational time of the same calculations on ORNL's current supercomputer Summit \cite{summit}. Each calculation was performed on one Summit node, utilizing all 6 NVIDIA V100 GPUs on the node. 

Results for the FePt alloy are shown in Table \ref{table_times1}. Similar results have been obtained also for CuAu. Our DL approach demonstrates significant time reductions for both CuAu and FePt cases when NN models are used in place of DFT calculations. The first-principles LSMS calculations take about 303 wall-clock seconds on average on a Summit node, whereas the NN models predict the physical quantities in about one wall-clock second.

The training of the NN models takes about 4,000 to 5,000 wall-clock seconds on an NVIDIA V100 GPU depending on the number of quantities that the NN is trained on, which affects the complexity of the objective function to minimize.

\begin{table}[h]
\centering
  \begin{tabular}{ccc}
    \toprule
    \textbf{Computational approach} & \textbf{Compute resources} &
      \textbf{Wall-clock time(s)}  \\
      \midrule
    1 DFT calculation & 1 Summit node & 303.2 \\
   prediction with single-tasking NN & 1 NVIDIA-V100 GPU & 0.9 \\
   prediction with multitasking NN & 1 NVIDIA-V100 GPU & 1.1\\
   \hline
   $\sim 10^6$ DFT calculations  & 1 Summit node & $ 3.03\cdot 10^8$ \\
   32,000 DFT calc. + training NN + $10^6$ NN pred.& 1 Summit node  & $1.10\cdot 10^6$ \\   
    \bottomrule
  \end{tabular}
  \caption{FePt binary alloy - Average time in wall-clock seconds needed to estimate macroscopic physical properties on a random lattice configuration with first principles calculations, time for one single-tasking NN models evaluation, time for one multitasking NN evaluation, total wall-clock time to perform $10^6$ DFT calculations and total wall-clock time to perform 32,000 DFT calculations, train the NN and evaluate the total energy for $10^6$ configurations using the NN.}
  \label{table_times1}
\end{table}

{Table \ref{table_times1} compares the computational time to compute the total energy for $10^6$ configurations using the LSMS code, with the time needed for the scenario where the total energy is calculated just for $32,000$ configurations with LSMS to generate the dataset, train the NN model on this dataset, then use the NN model to predict the total energy for $10^6$ alloy configurations. This comparison is of relevance for the possible use of surrogate DL models in Monte Carlo simulations to predict the thermodynamic properties of a solid solution alloy, in which the number of Monte Carlo samples required would at least be of the order of $10^6$ for sufficient sampling. In this case, it is clear that using DL surrogate models for accurate predictions of total energy can significantly reduce the computational time needed to predict material properties that would be otherwise unattainable given limited computational resources.}

\subsection{Comparison between statistical models for predictive performance}

In this section we compare the performance of NN models trained on a single task and the performance of NN models trained on multiple tasks. The output of DFT calculations are used as reference values. Therefore, the predictive performance of NN models measures the deviation of values predicted with DL approaches from the first principles calculations that are treated as exact. 

While the number of hidden layers (depth) of the NN architectures used in single-tasking and multitasking is different, the number of neurons in a hidden layers (width) is the same for both NN architectures in our study, as opposed to the analysis in \cite{Caruana} where NN models with different number of neurons per hidden layers were considered. Fixing the width of both NN architectures increases the generality and accuracy of the multitasking NN model, because the multitasking NN is forced to represent a joint function more compactly to predict all the quantities. This prevents the situation where multiple functions would be represented by different independent regions of an NN if the NN size is too large. 

{The root mean-squared error (RMSE), i.e., the Euclidean distance between the predictions and the values on the test set, is used to measure the accuracy of the NN model. To assess the predictive performance of the single-tasking and multitasking training methods, we measured the accuracy and the uncertainty of the model in the following manner: for every training method, the k-fold cross-validation results in $k$ NN models, each corresponds to an RMSE to indicate the accuracy of that particular NN model. The accuracy of the training method as a whole is then quantified by the mean value of the RMSE's calculated from all these $k=60$ predictions, and the uncertainty is quantified by the standard deviation of the mean of all these $k=60$ RMSE's.}

\begin{table}[h]
\centering
  \begin{tabular}{c|cc|cc}
    \toprule
    \multirow{2}{*}{\bf{{}Training method}} &
      \multicolumn{2}{c|}{\bf{RMSE - Total Energy} } &
      \multicolumn{2}{c}{\bf{RMSE - Charge Density} } \\
      & {Mean} & {Std. dev.} & {Mean} & {Std. dev.} \\
      \midrule
    multitasking, EC & $3.87 \times 10^{-3}$ & $1.96\times 10^{-3}$ & $8.47 \times 10^{-3}$ & $1.05\times 10^{-3}$ \\
    single-tasking, E & $6.65\times 10^{-3}$ & $3.15 \times 10^{-3}$ & - & - \\
    single-tasking, C & - & - & $3.27\times 10^{-3}$ & $4.28\times 10^{-4}$ \\
    \bottomrule
  \end{tabular}
  \caption{Training of neural network (NN) models to predict physical properties of CuAu alloy - mean values and standard deviations of RMSE are obtained over 60 runs. Each run performs the training of multitasking and single-tasking NN to predict the total energy and charge density. The name of each training method refers to whether it is a multitasking or a single-tasking NN model. The capital letters are associated with the quantities predicted (E stands for total energy, C stands for charge density).}
  \label{cuau_table}
\end{table}

Tables \ref{cuau_table} and \ref{fept_table} show the sample mean and the standard deviation of the RMSE for each physical quantity predicted by the NN models. 
The sample mean of the RMSE provides a point-wise estimation of the predictive power of the model, whereas the standard deviation quantifies the reliability of the RMSE in effectively describing the predictive performance of the model.
{Statistical models with a lower RMSE are to be interpreted as more accurate, and smaller values of the standard deviation are to be interpreted as a more reliable prediction of the model. In fact, a low standard deviation for the RMSE means that the NN models' predictions are similar across different data splitting. This in turn indicates that the epistemic uncertainty of the models is smaller. The relevant features that characterize the dataset are well captured, and the model prediction is thus more stable.}

\begin{table}[h]
\centering
  \begin{tabular}{c|cc|cc|cc}
    \toprule
    \multirow{2}{*}{\bf{Training method}} &
      \multicolumn{2}{c|}{\bf{RMSE - Total Energy} } &
      \multicolumn{2}{c|}{\bf{RMSE - Charge Density} } &
      \multicolumn{2}{c}{\bf{RMSE - Mag. Moment} } \\
      & {Mean} & {Std. dev.} & {Mean} & {Std. dev.} & {Mean} & {Std. dev.} \\
      \midrule
    multitasking, ECM & $5.97\times 10^{-3}$ & $8.85 \times 10^{-4}$ & $6.12\times 10^{-2}$ & $5.65\times 10^{-3}$ & $2.02\times 10^{-2}$ & $1.88\times 10^{-3}$ \\
    multitasking, EC & $6.22\times 10^{-3}$ & $1.18\times 10^{-3}$ & $5.75\times 10^{-2}$ & $3.25\times 10^{-3}$ & -& - \\
    multitasking, EM & $6.58\times 10^{-3}$ & $3.32\times 10^{-4}$ & - & - & $1.82\times 10^{-2}$ & $2.46\times 10^{-3}$ \\
    multitasking, CM & - & - & $6.18\times 10^{-2}$ & $3.90\times 10^{-3}$ & $2.01\times 10^{-2}$ & $1.83\times 10^{-3}$  \\
    single-tasking, E & $1.22\times 10^{-2}$ & $2.15 \times 10^{-3}$ & - & - & - & - \\
    single-tasking, C & - & - & $5.36\times 10^{-2}$ & $3.19\times 10^{-3}$ & - & - \\
    single-tasking, M & - & - & - & - & $1.55\times 10^{-2}$ & $1.76\times 10^{-3}$ \\
    \bottomrule
  \end{tabular}
  \caption{Training of neural network (NN) models to predict physical properties for FePt alloy - mean values and standard deviations of RMSE are obtained over 60 runs. Each run performs the training of multitasking or single-tasking NN to predict total energy, charge density and magnetic moment. The name of each training method refers to whether it is a multitasking or a single-tasking NN model. The capital letters are associated with the quantities predicted (E stands for total energy, C stands for charge density and M stands for magnetic moment).}
  \label{fept_table}
\end{table}

Since the number of nodes in a hidden layer of the NN models is the same, the RMSE for multitasking NN models may be higher compared to single-tasking NN models, because the same amount of computational resources is shared among multiple quantities to be predicted. 
In traditional ML studies, the performance of a model would be measured by its level of generalizability in retaining relevant features of the dataset, thereby making an accurate prediction for inputs outside the training set. This translates into the ability to avoid overfitting, which is usually diagnosed by a training error much lower than the validation error. 
Here, we adopt a more generic concept of predictive performance of a model: our entire procedure leads to a more generalizeable model, regardless of the specific dataset used for the training. 
Therefore, we do not only look at the RMSE, hence the accuracy, of a single model. In our case, we evaluate the effectiveness of the joint training procedure in stabilizing the predictions. In this regard, the standard deviation of the RMSE's of an ensemble of NN models is more informative. It can be viewed as a metric for quantifying the uncertainty associated with the predictions. The more stable a prediction is, the more likely the accuracy attained by the model is independent of the specific portion of dataset used for training the model itself. Therefore, more robust and stable models should lead to smaller standard deviation of the RMSE. If the joint training causes the RMSE to increase and the standard deviation to increase at the same time, this means that the quantities to be predicted are weakly correlated.  

\begin{figure}[h]
    \centering
    \includegraphics[width=0.45\textwidth]{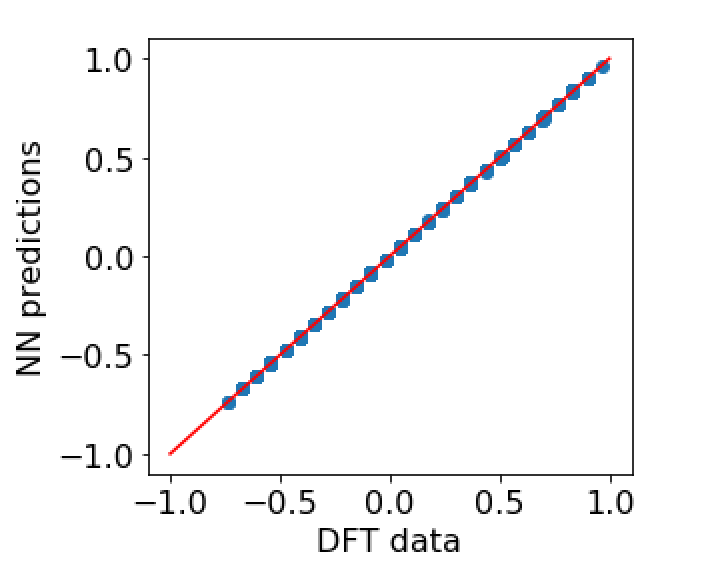}
    \includegraphics[width=0.45\textwidth]{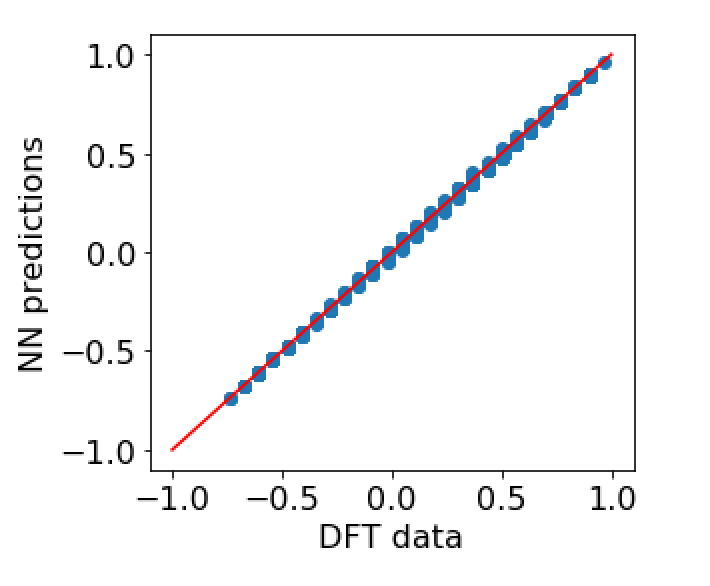}
    \caption{CuAu test case - scatter plot for standardized total energy predicted with NN against standardized DFT calculation using multitasking NN to predict total energy and charge density (left) and single tasking NN to predict energy (right).}
    \label{scatterplot_test1}
\end{figure}

With regard to the CuAu test case, a more noticeable reduction of the loss function (Equation \ref{global_loss}) for the total energy on the testing set is noticed when total energy and charge density are predicted simultaneously compared to the single-tasking NN that predicts the total energy alone. Results in Table \ref{cuau_table} show a decrease of the RMSE along with a significant decrease in the standard deviation for the total energy when it is predicted with multitasking NN models. The decrease of the standard deviation of the RMSE for the total energy means that the predictions are less prone to fluctuations, making the outcome of the predictive model more stable and reliable. 
This suggests that total energy and charge density are correlated, and the correlation allows the total energy to be better predicted because the additional information provided by the charge density acts as a relevant physical constraint. Although the mean value and standard deviation of RMSE for the charge density does not necessarily improve with the joint training, we remind the reader that the charge density is just an auxiliary quantity added as a physical constraint to improve the prediction of the total energy, which is the actual quantity we aim to better estimate. {The scatter plots in Figure \ref{scatterplot_test1} confirms that the predictions obtained with the multitasking NN better agree with the DFT reference than the single-tasking NN.}
  

This observation is further confirmed by the outcome of the FePt test case with the same joint training that involves total energy and charge density (see Table \ref{fept_table}). In addition, the magnetic moment can be also included in the multitasking learning as a second physical constraint because FePt is a magnetic alloy. Indeed, the total energy predicted by the NN model that simultaneously predicts total energy and charge density is also more precise than that predicted by the single-tasking NN. 
 
Results in Table \ref{fept_table} show that adding the magnetic moment as a physical constraint improves the predictive performance of multitasking NN models over the single-tasking NN predicting total energy only. The smaller RMSE for the total energy suggests that the training benefits from the inclusion of the magnetic moment as a constraint. This is also validated by the smaller standard deviation of the RMSE for the total energy when the multitasking NN is used instead of the single-tasking NN, meaning that the predictions of the total energy generated by the joint training are more stable. 

The results obtained by our DL approach suggest that the correlation between total energy and magnetic moment is stronger than the correlation between total energy and charge density. In fact, the combination of total energy and magnetic moment in the multitasking learning lowers the standard deviation of the RMSE for the total energy more effectively than with the charge density, meaning that the prediction of the total energy is more efficiently stabilized. Our observation of the stronger correlation between total energy and magnetic moment in ferromagnetic materials is also supported by previous DFT calculations and laboratory experiments \cite{pinski, hou, marshal}.

\begin{figure}[h]
    \centering
    \includegraphics[width=0.45\textwidth]{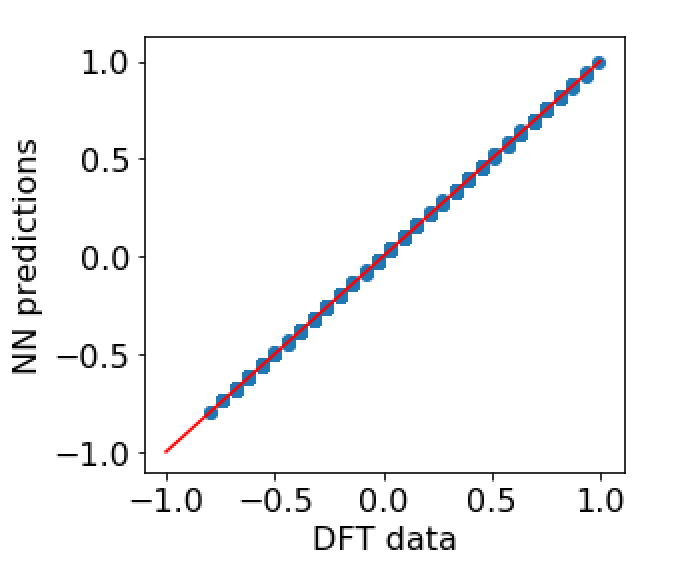}
    \includegraphics[width=0.45\textwidth]{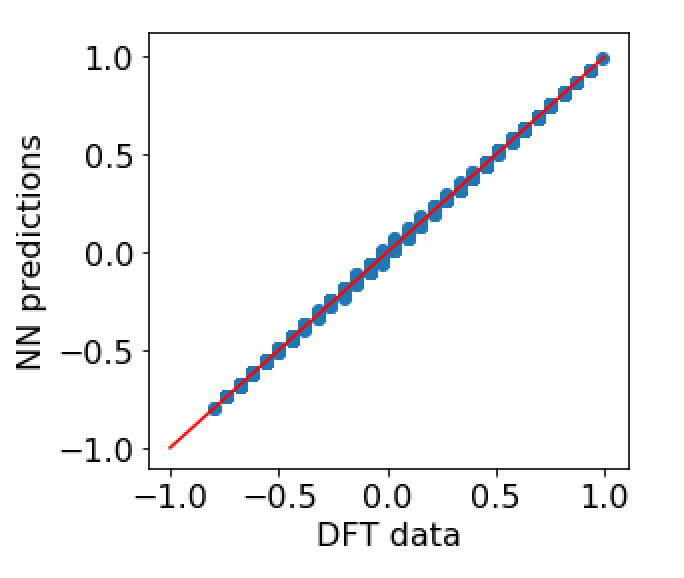}
    \caption{FePt test case - scatter plot for standardized total energy predicted with NN against standardized DFT calculation using multitasking NN to predict total energy, charge density and magnetic moment (left) and single tasking NN to predict energy (right).}
    \label{scatterplot_test2}
\end{figure}

Similar to the CuAu case, the mean values and standard deviations of RMSE for the charge density and magnetic moment do not necessarily improve with the joint training. Again, it is because the charge density and the magnetic moment are just auxiliary quantities added as physical constraints to improve the prediction of the total energy. The scatter plots in Figure \ref{scatterplot_test2} for the FePt test case also confirms that the predictions obtained with the multitasking NN better fits the data distribution than the single-tasking NN.

{While the joint training did not result in any appreciable improvement for the prediction of charge densities and magnetic moments, it should be noted that the total number of nodes per layer in our experiment remained constant. As the final, task specific, hidden layer has significantly reduced dimensionality compared to the single-tasking case, the capability to represent the spatially distributed quantities is reduced more severely than is the case for the total energy that can be well represented with the reduced number of final nodes provided. This observation suggests that this split in the final layer is another hyperparameter that can be optimized if a different focus is required.}

\section{Conclusions and future developments}
\label{conclusions}
In this paper we presented an approach to improve the reliability of NN models when they are used as a replacement of first principles calculations to calculate macroscopic material properties of binary alloys. The approach resorts to a multitasking NN model that is jointly trained on multiple quantities to reduce the uncertainty. Each predicted quantity acts as a physical constraint on the other quantities so that the reliability of the prediction is improved by forcing the NN models to learn the information for all the physical quantities.

Numerical experiments for CuAu and FePt alloys are presented where material properties are estimated for systems of 32 atoms placed on FCC and BCC lattices, respectively. The use of NN models effectively reduces the time needed to predict the material properties by a factor of hundreds of wall-clock seconds compared to direct calculations. 
Improvements in the stability of the prediction have been obtained for both test cases. The correlation of charge density and magnetic moment with the energy is found to stabilize the predictive performance in multitasking NN models for the prediction of the total energy.

Although the numerical experiments presented here clearly show that multitasking can use strong correlations between physical properties to reduce the uncertainty associated with the DL predictions, imposing too many constraints may be counterproductive. In fact, an excessively demanding predictive task that requires the joint training to simultaneously predict too many quantities can lead to a decrease of the predictive power. Therefore, the inclusion of the physical constraints based on correlations between physical properties must be consciously handled. 

We would also like to generalize the study by using a convolutional neural network (CNN) to capture local interactions between atoms. The use of CNN could reduce the computational cost to train the DL model compared to fully connected neural networks. Additionally, this will also make the DL model independent of the lattice size. Another direction we aim to pursue to generalize our approach is to allow atomic and volume relaxations to produce more accurate estimates that match the chemical compositions of materials in nature. {As noted in Section \ref{physics}, our current models were trained to a fixed lattice constant and volume for all chemical compositions in the binary alloys. More realistic models should be able to account for the atomic and volume relaxation for different chemical compositions. This would require including, as additional features in the input data, the atomic positions which would automatically provide the lattice constant and total volume. Additionally, the training dataset should be extended to include multiple volumes at different compositions to correctly capture the volume dependence of the total energy. The increase in dimensionality in the input data would require the collection of an even larger dataset, as well as a longer time to train the NN models.}

As anticipated earlier, the use of surrogate DL models to quickly estimate physical properties enables the studies of statistical mechanics for system sizes that otherwise would not be feasible given current computing limitations to perform DFT calculations. Therefore, DL model can replace DFT calculations for the estimate of the total energy at each acceptance-rejection step of a Monte Carlo algorithm.
When the volume of the data to handle is too large, an effective reduction of the size of the training set could be achieved either by constructing NN models that automatically recognize crystal symmetries or by using reinforcement learning \cite{prog_NN} to iteratively update the set-up of the NN based on new data generated on the fly. 

\section*{Acknowledgements}
This work was supported in part by the Office of Science of the Department of Energy and by the Laboratory Directed Research and Development (LDRD) Program of Oak Ridge National Laboratory. This work used resources of the Oak Ridge Leadership Computing Facility, which is supported by the Office of Science of the U.S. Department of Energy under Contract No. DE-AC05-00OR22725. Y. W. Li was supported by the LDRD Program of Los Alamos National Laboratory (LANL) under project number 20190005DR. LANL is operated by Triad National Security, LLC, for the National Nuclear Security Administration of U.S. Department of Energy (Contract No. 89233218CNA000001). K. Barros acknowledges support from the Center of Materials Theory as a part of the Computational Materials Science (CMS) program, funded by the U.S. Department of Energy, Office of Science, Basic Energy Sciences, Materials Sciences and Engineering Division. This document number is LA-UR-20-27481.

\section*{References}

\bibliographystyle{vancouver}
\bibliography{bibliography}

\end{document}